\documentclass[conference]{IEEEtran}
\IEEEoverridecommandlockouts
\usepackage{cite}
\usepackage{amsmath,amssymb,amsfonts}
\usepackage{algorithmic}
\usepackage{graphicx}
\usepackage{textcomp}

\usepackage{hyperref}
\hypersetup{
    colorlinks=true,
    linkcolor=blue,
    filecolor=magenta,      
    urlcolor=cyan,
    pdftitle={Overleaf Example},
    pdfpagemode=FullScreen,
    }

\usepackage{xcolor}
\def\BibTeX{{\rm B\kern-.05em{\sc i\kern-.025em b}\kern-.08em
    T\kern-.1667em\lower.7ex\hbox{E}\kern-.125emX}}

\begin{document}

\title{The Application of Quantum Fourier Transform in Cosmic Microwave Background Data Analysis}

\author{\IEEEauthorblockN{Farida Farsian}
\IEEEauthorblockA{\textit{OACT, INAF} \\
Via S. Sofia 78, Catania, Italy \\
farida.farsian@inaf.it}
\and
\IEEEauthorblockN{Tiziana Trombetti, Carlo Burigana}
\IEEEauthorblockA{\textit{IRA, INAF\textsuperscript{1}, INFN\textsuperscript{2}} \\
\textsuperscript{1}Via Pietro Gobetti 101, Bologna, Italy \\
\textsuperscript{2}Via Irnerio 46, Bologna, Italy \\
tiziana.trombetti@inaf.it, carlo.burigana@inaf.it}\\
\and
\IEEEauthorblockN{Francesco Schillir\'o}
\IEEEauthorblockA{\textit{OACT, INAF} \\
Via S. Sofia 78, Catania, Italy \\
francesco.schilliro@inaf.it}
\and
\IEEEauthorblockN{Andrea Bulgarelli}
\IEEEauthorblockA{\textit{OAS, INAF} \\
Via Pietro Gobetti 93/3, Bologna, Italy \\
andrea.bulgarelli@inaf.it}
\and
\IEEEauthorblockN{Vincenzo Cardone}
\IEEEauthorblockA{\textit{OAR, INAF} \\
Via Frascati 33, Monteporzio Catone, Italy \\
vincenzo.cardone@inaf.it}
\and
\IEEEauthorblockN{Luca Cappelli}
\IEEEauthorblockA{\textit{OATs, INAF} \\
Via Tiepolo 11, Trieste, Italy \\
luca.cappelli@inaf.it}
\and
\IEEEauthorblockN{Massimo Meneghetti}
\IEEEauthorblockA{\textit{OAS, INAF} \\
Via Pietro Gobetti 93/3, Bologna, Italy \\
massimo.meneghetti@inaf.it}
\and
\IEEEauthorblockN{Giuseppe Murante}
\IEEEauthorblockA{\textit{OATs, INAF} \\
Via Tiepolo 11, Trieste, Italy \\
giuseppe.murante@inaf.it}
\and
\IEEEauthorblockN{Alessandro Rizzo}
\IEEEauthorblockA{\textit{OACT, INAF} \\
Via S. Sofia 78, Catania, Italy \\
alessandro.rizzo@inaf.it}
\and
\IEEEauthorblockN{Giuseppe Sarracino, Irene Graziotti}
\IEEEauthorblockA{\textit{OACN, INAF} \\
Via Moiariello 16, Napoli, Italy \\
giuseppe.sarracino@inaf.it}
\and
\IEEEauthorblockN{Roberto Scaramella, Vincenzo Testa}
\IEEEauthorblockA{\textit{OAR, INAF} \\
Via Frascati 33, Monteporzio Catone, Italy \\
roberto.scaramella@inaf.it, vincenzo.testa@inaf.it}}

\maketitle

\begin{abstract} 
The Cosmic Microwave Background (CMB) data analysis and the map-making process rely heavily on the use of spherical harmonics. For suitable pixelizations of the sphere, the (forward and inverse) Fourier transform plays a crucial role in computing all-sky map from spherical harmonic expansion coefficients -- or from angular power spectrum -- and vice versa. While the Fast Fourier Transform (FFT) is traditionally employed in these computations, the Quantum Fourier Transform (QFT) offers a theoretical advantage in terms of computational efficiency and potential speedup. In this work, we study the potential advantage of using the QFT in this context by exploring the substitution of the FFT with the QFT within the \textit{healpy} package. Performance evaluations are conducted using the Aer simulator. Our results indicate that QFT exhibits potential advantages over FFT that are particularly relevant at high-resolution. However, classical-to-quantum data encoding overhead represents a limitation to current efficiency. In this work, we adopted amplitude encoding, due to its efficiency on encoding maximum data to minimum number of qubits. 
We identify data encoding as a potential significant bottleneck and discuss its impact on quantum speedup. Future improvements in quantum encoding strategies and algorithmic optimizations could further enhance the feasibility of QFT in CMB data analysis.
\end{abstract}


\section{Introduction}
\label{intro}
In recent years, advancements in quantum computing hardware and its increasing accessibility have provided a foundation for exploring its potential across various scientific domains. Quantum computers offer remarkable advantages in terms of speed and computational efficiency, particularly in areas such as prime number factorization \cite{365700}, quantum system simulations \cite{Lloyd1996UniversalQS}, and solving linear systems of equations \cite{Harrow_2009}. These rapid developments have opened new pathways for addressing complex computational challenges that were previously infeasible with classical methods.

In astrophysics and cosmology, many computational problems demand significant resources, necessitating the exploration of novel methodologies to improve efficiency \cite{doi:10.1063/1.5116791}. One of the fundamental pillars of modern cosmology is the study of the Cosmic Microwave Background (CMB), the residual thermal radiation from the early universe. A particularly demanding computational task in this field is CMB map-making, where observational data is transformed into high-resolution sky maps and vice versa \cite{Challinor2017}. Given the importance of Fourier transforms in this process, and the potential computational speedups offered by quantum algorithms \cite{Bracewell2000}, this paper investigates the feasibility of replacing the classical Fast Fourier Transform (FFT) with the Quantum Fourier Transform (QFT) in CMB data analysis. By assessing the performance of QFT within this context, we aim to evaluate its potential advantages and identify key bottlenecks. 

The structure of this paper is as follows: 
Section \ref{cmb} provides an overview of the CMB, discussing its significance and key properties. In Section \ref{map}, we describe the process of CMB map-making and its associated challenges. Section \ref{qft} introduces the QFT, explaining its theoretical foundations and potential advantages. Section \ref{method} describes the implementation of QFT for CMB data processing, outlining the methodology and computational framework. In Section \ref{res}, we present the experimental setup and results, analyzing the performance and accuracy of QFT compared to FFT. Finally, Section \ref{conclu} discusses the implications of our findings and concludes the paper, highlighting potential future research directions.

\section{Cosmic Microwave Background}
\label{cmb}
The Cosmic Microwave Background (CMB) provides a snapshot of the cosmos approximately 380,000 years after the Big Bang \cite{Peebles1970}. This radiation, first discovered by Penzias and Wilson in 1965 \cite{Penzias1965}, is a relic of the recombination epoch when the universe cooled sufficiently for protons and electrons to combine into neutral hydrogen, allowing photons to travel freely. The CMB is observed today as a nearly uniform blackbody radiation with a temperature of 2.72548 K \cite{2009ApJ...707..916F}, but with small anisotropies encoding essential cosmological information.

The importance of the CMB in cosmology cannot be overstated. It serves as one of the primary observational pillars for the Big Bang theory \cite{Gamow1948} and provides crucial constraints on fundamental cosmological parameters. The temperature fluctuations within the CMB, on the order of $\mu$K, arise from primordial density perturbations that later evolved into the large-scale structure of the universe, including galaxies and clusters. These anisotropies are key to testing theories of cosmic inflation \cite{Guth1981}, determining the universe's composition (e.g., dark matter and dark energy), and measuring parameters such as the Hubble constant, baryon density, and spectral index of primordial perturbations.

Detailed observations of the CMB, particularly through experiments like COBE \cite{Smoot1992}, WMAP \cite{Bennett2003}, and \textit{Planck} \cite{Planck2018}, have refined our understanding of the early universe with unprecedented precision. The CMB power spectrum, derived from its temperature and polarization anisotropies, provides insights into the universe’s geometry, confirming that it is nearly spatially flat. Moreover, secondary effects such as the Sunyaev-Zeldovich effect \cite{Sunyaev1970} and gravitational lensing of the CMB \cite{Lewis2006} further enhance its role in probing the evolution of cosmic structure.

\section{CMB Map-Making and Its Challenges}
\label{map}
The reconstruction of CMB anisotropy maps from observational data is a fundamental task in modern cosmology. The process involves collecting data in the form of sky temperature fluctuations from satellite, balloon-borne, or ground-based experiments and converting them into a systematic representation on a celestial sphere. This transformation is critical for extracting cosmological information, requiring accurate modeling of instrumental effects, noise filtering, and deconvolution techniques \cite{dodelson2003modern, hu2002cosmic}.

One of the principal challenges in CMB map-making is the presence of systematic effects, including beam asymmetries, scanning strategies, correlated noise, and foreground contamination. These factors necessitate sophisticated algorithms capable of disentangling true CMB fluctuations from instrumental artifacts and astrophysical foregrounds. The map-making process typically begins with time-ordered data (TOD) recorded by detectors scanning the sky, which must be projected onto a pixelized sphere using methods such as maximum-likelihood estimators \cite{tegmark1997maximum} or destriping techniques \cite{mitra2004destriping}.

\subsection{The Role of Spherical Harmonics and Fourier Transforms in CMB Analysis}
Since the CMB temperature fluctuations are defined on a sphere, the most natural basis for their analysis is the spherical harmonics, \( Y_{\ell m}(	\theta, \phi) \). Any observed CMB temperature map, \( T(\theta, \phi) \), can be expanded as:
\begin{equation}
    T(	\theta, \phi) = \sum_{\ell=0}^{\infty} \sum_{m=-\ell}^{\ell} a_{\ell m} Y_{\ell m}(	\theta, \phi)
    \label{eq1}
\end{equation}

\noindent
where the spherical harmonic coefficients \( a_{\ell m} \) encode the statistical properties of the CMB anisotropies. These coefficients are crucial for estimating the angular power spectrum, \( C_\ell = \langle |a_{\ell m}|^2 \rangle \), which provides direct insight into the primordial density perturbations and fundamental cosmological parameters \cite{Peebles1970}.

In practical map-making and power spectrum estimation, the transformation from real-space temperature maps to the harmonic space (where statistical isotropy simplifies the analysis) requires an efficient implementation of the spherical harmonic transform. This transformation is mathematically analogous to the Fourier transform on the sphere, often computed using the Fast Spherical Harmonic Transform (FSHT) \cite{driscoll1994computing}. The FSHT serves as the spherical equivalent of the FFT in Euclidean space, enabling computationally efficient conversions between pixel-space and harmonic-space representations. For high-resolution datasets, the computational cost of the FSHT scales as \( O(N^{3/2}) \), where \( N \) is the number of pixels, making it a computational bottleneck for large-scale CMB experiments \cite{healy1998fast}.

\subsection{The map2alm Function in Healpy: Spherical Harmonic Decomposition of CMB Maps}
The function \textit{map2alm} in the Healpy package \cite{Zonca_2015} is a fundamental tool for analyzing CMB maps in harmonic space. It performs the spherical harmonic transform (SHT), which decomposes a pixelized sky map into a series of spherical harmonic coefficients, providing a representation that is particularly useful for statistical and cosmological analyses. This transformation is crucial for extracting anisotropies in the CMB and comparing observational data with theoretical models.

\subsubsection{Mathematical Formalism}
The \textit{map2alm} function computes the spherical harmonic coefficients \( a_{\ell m} \) from a discretized map of the CMB temperature anisotropies, \( T(\theta, \phi) \), using Equation \eqref{eq1}, where the expansion coefficients \( a_{\ell m} \) are given by the inverse transform:

\[a_{\ell m} = \int_{0}^{2\pi} \int_{0}^{\pi} T(\theta, \phi) Y_{\ell m}^{*}(	\theta, \phi) \sin \theta \ d\theta \ d\phi.\]

\noindent
Here, \(\ell\) is the multipole moment, which characterizes the angular scale of the spherical harmonic mode, and \(m\) is the azimuthal index, which determines the longitudinal variation of the mode. Numerical approximations to these integrals are performed using efficient quadrature techniques adapted to the HEALPix (Hierarchical Equal Area isoLatitude Pixelation) scheme  \cite{Gorski_2005}, which ensures a balance between resolution and computational efficiency.

\subsubsection{Computational Implementation and Performance}
The \textit{map2alm} function employs fast algorithms based on recurrence relations for computing associated Legendre polynomials, which are central to evaluating \( Y_{\ell m}(\theta, \phi) \). The efficiency of \textit{map2alm} is enhanced by leveraging symmetries in the HEALPix grid, enabling a computational complexity scaling approximately as \( \mathcal{O}(N_{\text{pix}}^{1.5}) \), where \( N_{\text{pix}} \) is the number of pixels in the input map. This scaling is discussed in \cite{Healpix_2004}.

Furthermore, this formalism allows for the computation of polarization components by extending the decomposition to Stokes parameters \( Q \) and \( U \), yielding the harmonic coefficients \( E_{\ell m} \) and \( B_{\ell m} \) through:
\[E_{\ell m} \pm i B_{\ell m} = \int_{0}^{2\pi} \int_{0}^{\pi} (Q \pm iU) {}_2Y_{\ell m}^{*} \sin \theta \ d\theta \ d\phi,\]

\noindent
where \( {}_2Y_{\ell m} \) are the spin-2 spherical harmonics. This capability is essential for separating the scalar \( E \)-mode and pseudo-scalar \( B \)-mode polarizations, which have distinct physical interpretations in CMB analysis \cite{Zaldarriaga_1997}.

\subsubsection{Importance in CMB Data Analysis}
The transformation from pixel space to harmonic space is a fundamental step in CMB data analysis. The power spectrum, \( C_\ell \), which characterizes the statistical properties of CMB fluctuations, is directly computed from the \( a_{\ell m} \) coefficients as:
\[C_\ell = \frac{1}{2\ell + 1} \sum_{m=-\ell}^{\ell} |a_{\ell m}|^2.\]

By applying \textit{map2alm}, one can efficiently extract these coefficients, facilitating the comparison of observed maps with theoretical predictions derived from cosmological models. Additionally, the function is widely employed in component separation, foreground subtraction, and beam deconvolution techniques, further enhancing its role in high-precision cosmological studies (see, for example, \cite{Planck_2016_XIII}).

\subsection{Challenges in Fourier Transform Applications}
While the FFT is central to standard map-making pipelines, its application to CMB analysis is constrained by the need for spherical geometry rather than a simple Euclidean grid. Standard FFT algorithms assume periodic boundary conditions and Cartesian symmetry, whereas CMB maps are naturally defined on the two-dimensional sphere. Consequently, numerical techniques such as HEALPix-based pixelization schemes \cite{Healpix2005} and specialized algorithms like the spin-weighted spherical harmonic transform \cite{SpinTransform} are required to properly handle the data.

Moreover, the presence of missing data, partial sky coverage due to galactic foreground masking, and anisotropic noise introduces additional challenges. These issues complicate the direct computation of spherical harmonic coefficients and necessitate iterative techniques, such as constrained realizations or Monte Carlo-based power spectrum estimation methods. Additionally, the non-uniform distribution of observational noise across the sky can lead to mode-coupling, which must be accounted for in unbiased cosmological parameter estimation.

\subsection{Need for Alternative Approaches}
Given these computational requirements, exploring more efficient methods for performing Fourier-like transforms on the sphere is essential. Quantum algorithms, particularly the QFT, offer a potential pathway for accelerating these operations. If implemented efficiently, QFT could significantly reduce the computational cost of transforming CMB maps to harmonic space, providing a novel approach to handling large-scale cosmological datasets.


\section{Quantum Fourier Transform}
\label{qft}
Quantum Fourier Transform (QFT) is the quantum analogue of the classical Discrete Fourier Transform (DFT) and serves as a fundamental subroutine in various quantum algorithms \cite{nielsen2010quantum}. Unlike the classical Fourier Transform, which requires \(O(N^2)\) operations for an input of size \(N\), the Fast Fourier Transform (FFT) reduces this cost to \(O(N \log N)\). QFT, however, leverages quantum parallelism to achieve an asymptotic complexity of \(O(\log^2 N)\) \cite{cleve1998quantum}, making it highly efficient for large-scale problems given sufficient quantum resources.

Mathematically, QFT acts on a quantum state encoded in the computational basis as follows \cite{kaye2007introduction}:
\[ |x\rangle = \sum_{j=0}^{N-1} x_j |j\rangle \to \sum_{k=0}^{N-1} \tilde{x}_k |k\rangle, \]

\noindent
where \(\tilde{x}_k\) are the Fourier coefficients of the input amplitudes. The transformation is defined by the unitary matrix:
\[ U_{jk} = \frac{1}{\sqrt{N}} e^{2\pi i jk / N}. \]

\subsection{Steps of the Quantum Fourier Transform}
QFT is implemented through a sequence of Hadamard gates and controlled phase shifts \cite{barenco1995elementary}, following these key steps:

1. Hadamard Transformation: The first qubit undergoes a Hadamard gate, creating an equal superposition state.

2. Controlled Phase Rotations: Each subsequent qubit undergoes controlled phase shifts, where the phase shift applied to the \(j\)-th qubit depends on the state of the previous qubits. This introduces quantum interference necessary for Fourier transformation.

3. Bit Reversal Permutation: Due to the nature of QFT computation, the output qubits must be rearranged in bit-reversed order to obtain the correct Fourier coefficients.



\subsection{QFT in Quantum Computation}
QFT serves as the core of many quantum algorithms, including Shor’s factoring algorithm \cite{shor1997polynomial} and phase estimation techniques \cite{abrams1999quantum}. In the context of CMB analysis, its efficiency in processing large datasets suggests potential advantages over classical FFT methods, particularly in reducing computational complexity for large-scale transformations.

\section{Methodology} 
\label{method}

In this section, we describe the steps and tools used to implement the QFT as a replacement for the classical FFT routine in CMB data analysis.  

\subsection{CMB Map Processing} 

We generate synthetic CMB maps using the theoretical power spectrum computed with the \textit{CAMB} (Code for Anisotropies in the Microwave Background) software \cite{Lewis_2000}. CAMB is a widely used cosmological Boltzmann code that computes the evolution of cosmological perturbations and predicts theoretical power spectra for CMB temperature and polarization anisotropies.  

For this study, we adopt a \(\Lambda\)CDM  cosmological model with the following parameters: \\
- \( H_0 = 67.5 \): The Hubble constant in km/s/Mpc, describing the expansion rate of the universe.  \\
- \( \Omega_b h^2 = 0.022 \): The baryon density parameter, which quantifies the fraction of the total energy density contributed by ordinary matter.  \\
- \( \Omega_c h^2 = 0.122 \): The cold dark matter density parameter.  \\
- \( m_\nu = 0.06 \) eV: The sum of neutrino masses, influencing the evolution of large-scale structure.  \\
- \( \Omega_k = 0 \): The curvature parameter, assuming a spatially flat universe.  \\
- \( \tau = 0.06 \): The optical depth to reionization, affecting the amplitude of the CMB polarization signal.  \\
- \( A_s = 2 \times 10^{-9} \): The scalar amplitude of primordial fluctuations.  \\
- \( n_s = 0.965 \): The scalar spectral index, determining the scale dependence of primordial fluctuations.\\
Given the theoretical power spectrum, we generate synthetic CMB maps using the \textit{alm2map} function from the \textit{healpy} package. This function reconstructs a pixelized sky map from a set of spherical harmonic coefficients \( a_{\ell m} \), obtained from CAMB. Once the simulated CMB maps are generated, we proceed with spherical harmonic decomposition using the \textit{map2alm} function. This step is performed using either the standard FFT-based approach or the proposed QFT-based approach.  

\subsubsection{Quantum Fourier Transform Implementation}

We implement QFT using the \textit{Qiskit} framework. The procedure consists of the following steps: \\
- \textbf{Data Normalization}: The pixel values of the CMB map are normalized to ensure compatibility with quantum encoding constraints. \\
- \textbf{Quantum State Preparation}: Qubits are initialized to represent the input amplitudes using amplitude encoding. \\
- \textbf{QFT Execution}: The QFT is applied to the quantum circuit, transforming the input data from real space to harmonic space. \\
- \textbf{Measurement and Reconstruction}: The transformed coefficients are obtained by measuring the quantum state, allowing for comparison with the classical FFT results.  

We evaluate the performance of QFT on a local Qiskit AerSimulator as detailed in the subsequent sections.  

\subsubsection{Traditional FFT-Based Decomposition} 

For benchmarking, we use the \textit{map2alm} function in the \textit{healpy} package, which employs the FFT for spherical harmonic decomposition. The execution time of the classical FFT routine is recorded for direct comparison with the QFT-based approach.  

Since quantum circuits operate on input vectors of length \( 2^n \), the pixel data must be padded to the nearest power of two before applying QFT. In the following subsections, we describe the data encoding strategy and the padding technique required for quantum circuit compatibility.  

\subsection{Data Encoding methods}

Quantum data encoding is the process of mapping classical data into a quantum state to leverage quantum computational advantages. This step is fundamental as it determines how efficiently and accurately classical data can be represented and processed within a quantum system. Depending on the nature of the data and the quantum algorithm, different encoding strategies are used, each with trade-offs in terms of resource requirements, scalability, and expressivity. Here we explain data encoding method used in this work:

\textbf{Amplitude Encoding}
maps classical data into the amplitudes of a quantum state. For a normalized classical vector \(\vec{x} = (x_1, x_2, \ldots, x_{2^n})\), the corresponding quantum state is:
\[
|\psi\rangle = \sum_{i=0}^{2^n-1} x_i |i\rangle.
\]
This approach is highly efficient in terms of qubit usage, as it allows \(2^n\) data points to be encoded using only \(n\) qubits. However, preparing such states can be computationally expensive, as it often involves complex unitary transformations, especially for large datasets.
Given the large volume of classical data that needs to be encoded in this work, amplitude encoding is the most suitable approach due to its exponential compression of information, enabling efficient representation within a limited number of qubits.


\subsection{Padding}

Considering the type of data encoding, a crucial preprocessing step would be transforming the data into a format compatible with quantum circuits. Specifically, amplitude encoding requires input vectors whose length is a power of two.
To ensure compatibility, we employ periodic padding, a method in which the data is extended by repeating itself periodically until the required power-of-two length is reached. Given an initial dataset of size $N$, the next power-of-two size is determined as $2^{\lceil \log_2 N \rceil}$. The periodic repetition ensures that the extended dataset maintains the original structure while avoiding artifacts introduced by zero-padding or arbitrary extrapolations.
This approach is particularly beneficial in quantum computing applications, as it preserves the frequency components of the original data, avoiding artificial discontinuities that could distort results when applying the QFT. By ensuring that the input data conforms to the necessary quantum hardware constraints, periodic padding enables seamless integration of classical cosmological datasets into quantum computing frameworks.


\section{Results}
\label{res}
To evaluate the performance of QFT as a replacement for the classical FFT, we conducted tests across multiple HEALPix resolutions, varying the \textit{nside} parameter from 2 to 2048 which corresponds to  \(12\times \text{nside}^2\) pixels. However, due to hardware limitations on available quantum processors, we restricted our tests on real quantum hardware to \(\text{nside} = 2\), which corresponds to six qubits. Each test was executed 10 times to ensure statistical robustness. Additionally, to validate the correctness of QFT-based results, we compared the extracted power spectrum against the theoretical spectrum computed using CAMB and the classical FFT-based method. Since the power spectrum serves as a crucial diagnostic tool in cosmological analysis, this comparison provides insight into the reliability of QFT for astrophysical applications.

\subsection{Time performance comparison}

\begin{figure}
    \centering
    \includegraphics[width=0.99\linewidth]{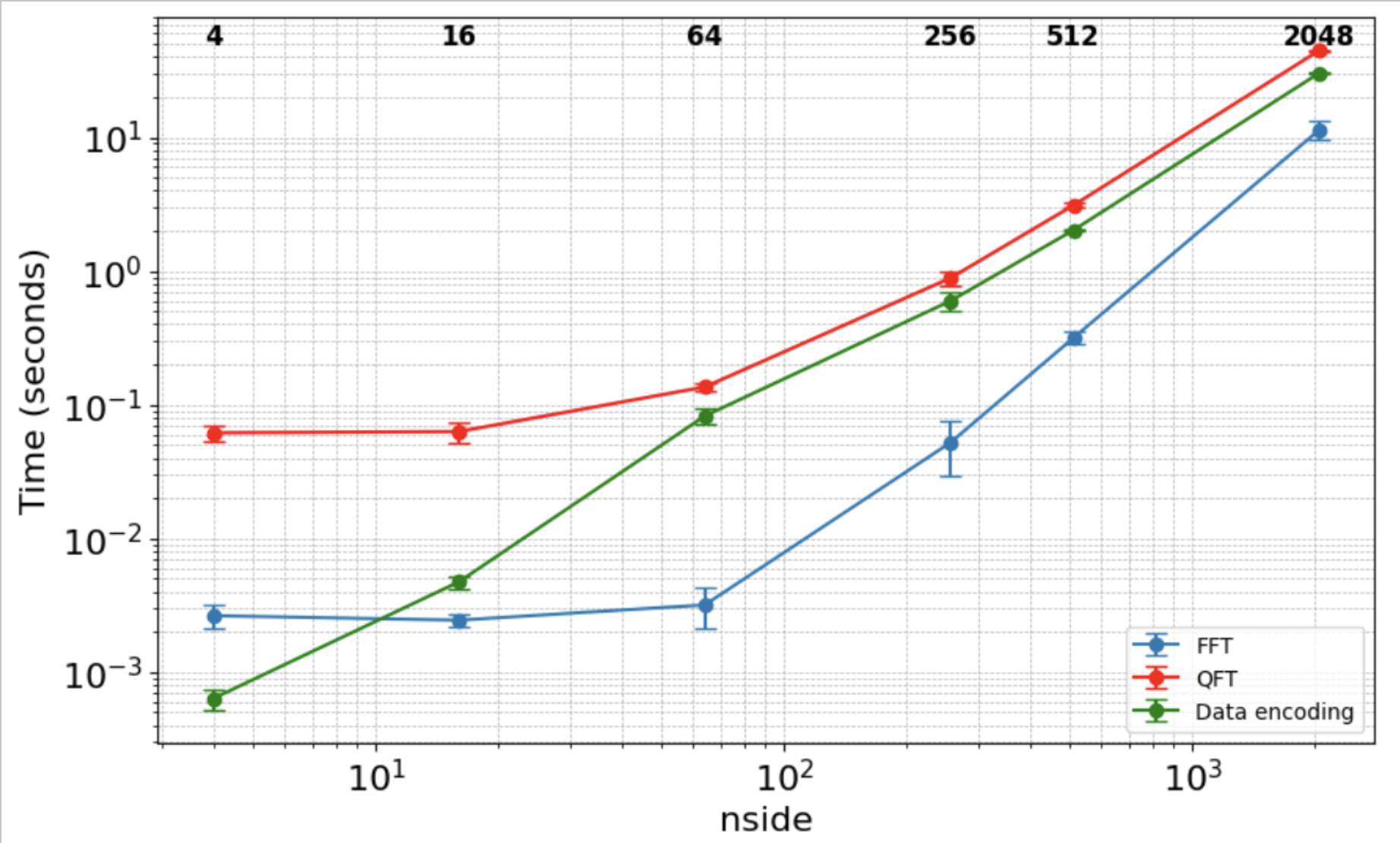}
    \includegraphics[width=0.99\linewidth]{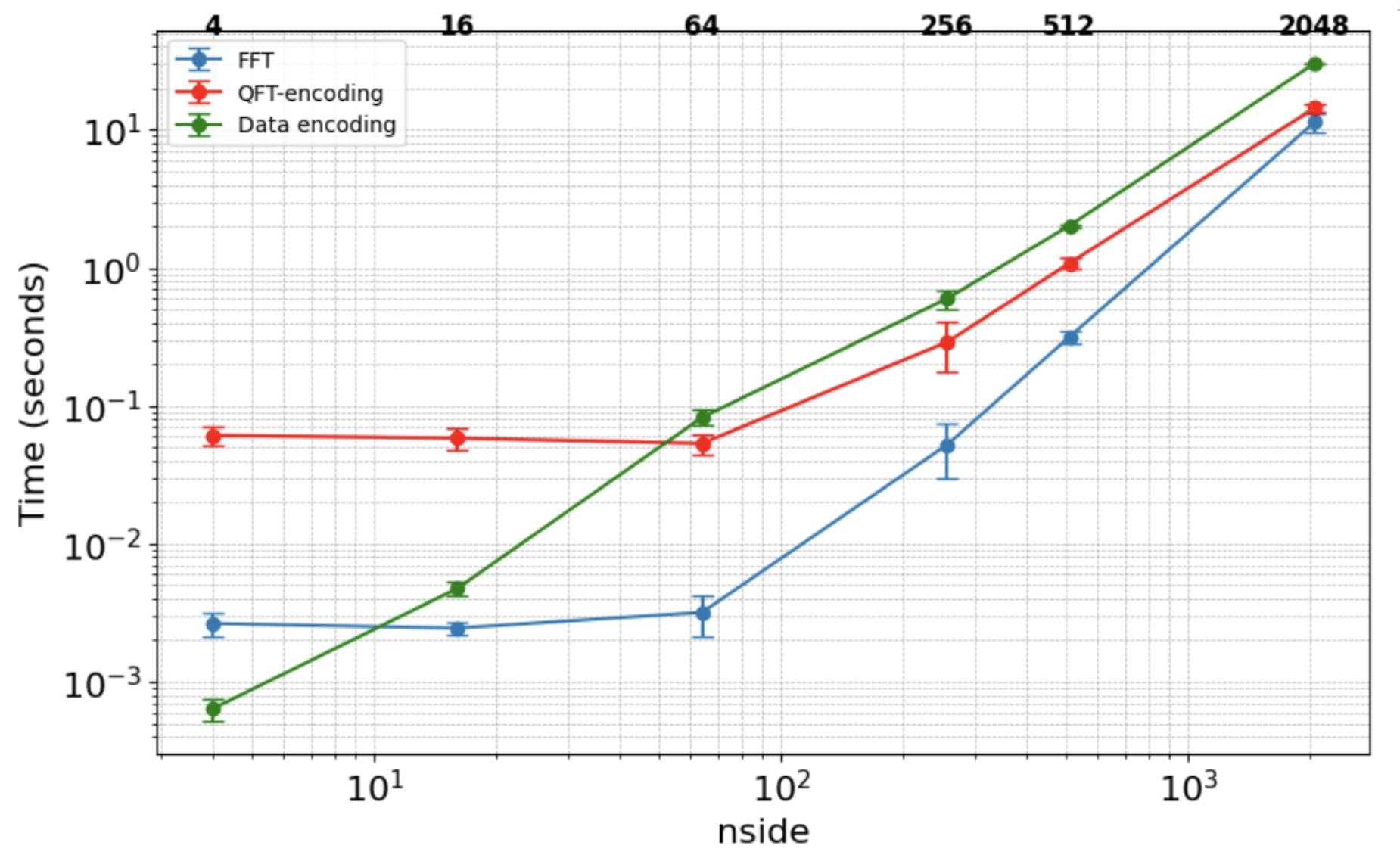}
    \caption{Comparison of FFT and QFT operation times as a function of HEALPix resolution parameter \textit{nside}. In the upper panel, the blue and red line represents the execution time of FFT and QFT. For the QFT, the timing includes circuit initialization, data encoding, quantum circuit execution on the Aer simulator, and measurement. The green line isolates the contribution of data encoding time. While in the lower panel, data encoding time is excluded from time measurement of red line, to show the effect of data encoding.}
    \label{fig1}
\end{figure}

Figure \ref{fig1} presents a comparative analysis of the execution times for the existing FFT subroutine in \textit{map2alm} function and the implemented  QFT subroutine to substitute the FFT in the same function. The blue curve represents the computational time required for the classical FFT in both panels, while the red curve corresponds to the total execution time of the QFT, including data encoding, quantum circuit initialization, quantum computation, and measurement, executed using the Aer simulator in the upper panel, while in the lower panel the data encoding time is subtracted from the red curve. The green curve isolates the contribution of data encoding, which in our case is performed using amplitude encoding. 
Each data point in Figure \ref{fig1} represents the mean execution time obtained from 10 independent runs to ensure statistical robustness. The error bars correspond to the standard deviation, providing a measure of the variability in execution time. 

At first glance, the results suggest that the FFT outperforms the QFT across all tested \textit{nside} values. However, a closer examination reveals that the primary bottleneck in the quantum approach arises from the data encoding process. This effect becomes particularly pronounced as \textit{nside} increases, meaning that for higher-resolution maps, the time required for encoding the classical data into quantum states dominates the total execution time. The scaling behavior of the QFT itself is not inherently limiting; rather, the inefficiencies introduced by amplitude encoding are responsible for the observed slowdown.
These results highlight a key challenge in leveraging quantum algorithms for CMB map-making: the classical-to-quantum data encoding overhead. The quantum advantage predicted in theoretical analyses is currently obscured by this overhead. However, if more efficient quantum data encoding techniques can be developed, the QFT could surpass the FFT in performance for high-resolution maps.

\begin{figure}
    \centering
    \includegraphics[width=0.99\linewidth]{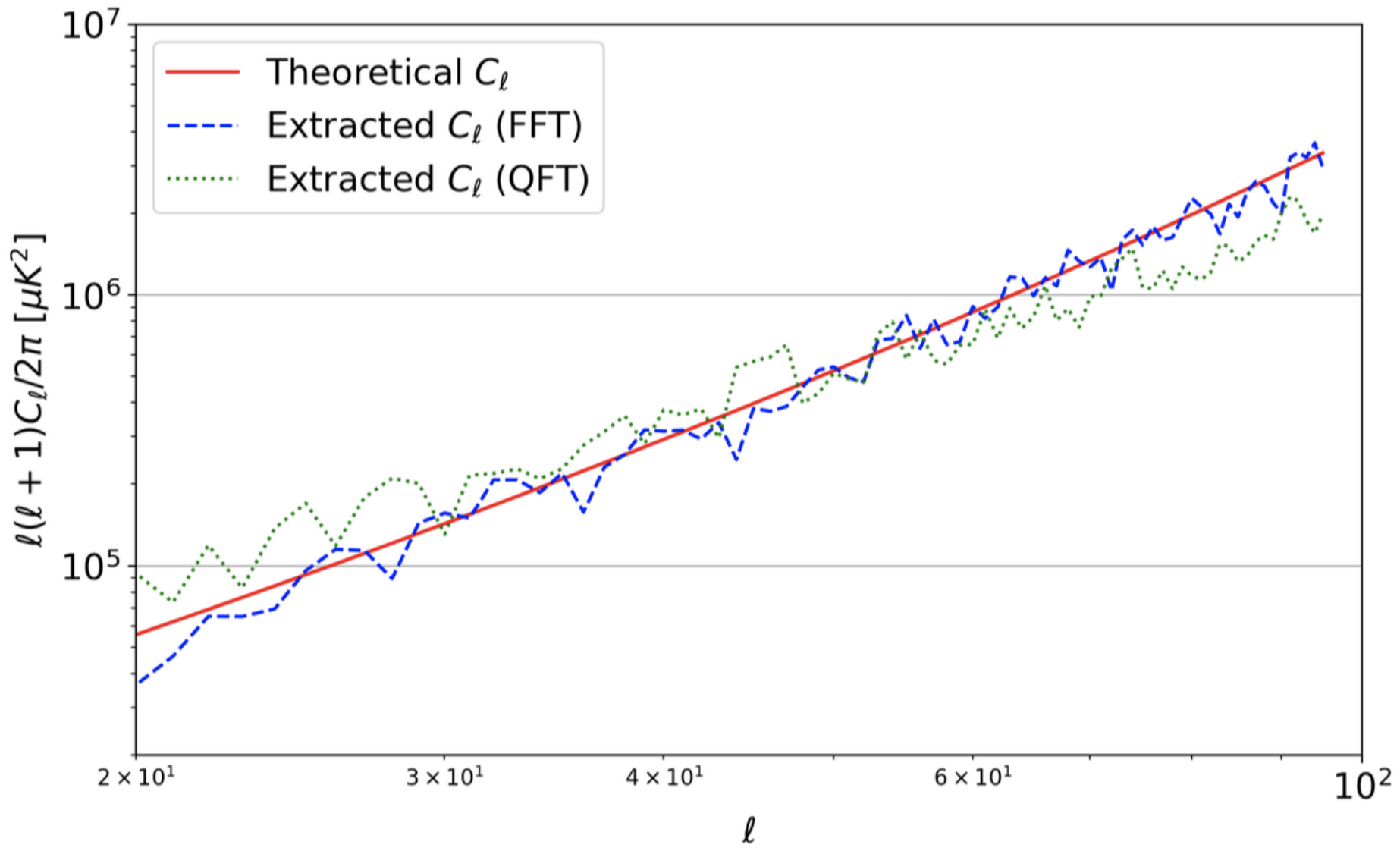}
    \caption{Comparison of the theoretical CMB power spectrum (\(C_\ell\)) with extracted spectra obtained using the FFT and QFT. The power spectra are computed from spherical harmonic coefficients (\(a_{\ell m}\)) derived using the \textit{map2alm} function, applied to CMB maps with \( \text{Nside} = 32 \).}
    \label{fig2}
\end{figure}  

\subsection{Power Spectrum Analysis} 
Figure \ref{fig2} presents a comparison between the theoretical CMB power spectrum (solid red line) and the spectra extracted using two different Fourier transform methods within the \textit{map2alm} function: the classical FFT (dashed blue line) and the QFT (dotted green line). The x-axis represents the multipole moment (\(\ell\)), which corresponds to different angular scales inversely proportional to the angular size, while the y-axis shows the power spectrum values, scaled as \(\ell(\ell+1)C_\ell/2\pi\) in units of \(\mu K^2\). The extracted power spectra exhibit fluctuations around the theoretical prediction due to numerical errors and resolution constraints (\(\text{nside} = 32\)). The QFT-based method shows increased deviations at lower \(\ell\) compared to FFT, which may be attributed to approximations introduced by the quantum algorithm. Conversely, at higher \(\ell\), a reduction in the power spectrum is observed, which may result from the additional pixels introduced into the map during the padding process.

In these experiments, we were able to push the analysis up to \(\text{nside} = 32\). Beyond this resolution, the computational complexity increased significantly due to the need to compute state vectors for QFT-based spherical harmonic coefficients (\(a_{\ell m}\)). This process involves matrix multiplications that impose high memory requirements, making it infeasible with the available computational resources.  

While the QFT implementation in this study was based on Qiskit, future work will explore alternative frameworks, such as PennyLane \cite{Pennylane} and Cirq \cite{Cirq}, to assess their efficiency and potential optimizations. Despite the limitations, the overall consistency between the power spectra extracted using QFT and the classical methods reinforces the reliability of our findings and highlights the necessity of further research into optimized quantum encoding strategies.  

\subsection{Execution on Real Quantum Hardware} 
The QFT-based algorithm was executed on an IBM quantum processor using the free-access quantum hardware, which provides six qubits (\(\text{nside} = 2\)). The total execution time recorded by the IBM platform was 3.17 seconds, encompassing all computational steps, including quantum circuit preparation, transpilation, data encoding, and QFT execution. However, due to platform constraints, it was not possible to isolate the execution time of the QFT alone.  

For comparison, the classical FFT-based routine completes the same operation in 0.0033 seconds for \(\text{nside} = 2\). While this suggests a significant computational overhead for QFT, this result is not indicative of the true efficiency of the quantum algorithm, as it includes all pre-processing and encoding steps. A fair comparison would require optimized quantum circuit execution and improved encoding strategies, which will be addressed in future work.

\section{Conclusion and future work}  
\label{conclu}
This study explored the feasibility of integrating the QFT into CMB data analysis, specifically in the spherical harmonic decomposition process used in the \textit{map2alm} function. Our results indicate that while QFT theoretically offers an exponential speedup over the classical FFT, the practical implementation is currently limited by the overhead of classical-to-quantum data encoding. Performance analysis demonstrated that the QFT execution time is dominated by the encoding step, rather than the quantum computation itself.

Despite these challenges, our findings suggest that QFT could become a viable alternative to FFT for high-resolution CMB maps if more efficient quantum encoding strategies are developed. Additionally, the accuracy of QFT-based power spectrum estimation is comparable to that of FFT, reinforcing its potential for future applications in CMB analysis. As quantum hardware continues to advance, addressing the bottlenecks associated with data encoding and noise mitigation will be crucial in realizing the full advantage of QFT in astrophysical data processing. 

Future research will focus on extending our analysis to higher-resolution maps, specifically pushing the \textit{nside} parameter to 4096 and beyond. This resolution was previously unattainable on our local machine due to memory limitations, and overcoming this constraint will enable a more comprehensive evaluation of QFT's performance at large scales. Additionally, we aim to execute our algorithm on real quantum hardware to assess its practical feasibility and compare performance metrics with those obtained from simulations.  

Another key direction is the development of a fully quantum end-to-end pipeline by incorporating the inverse QFT into the \textit{alm2map} function. This extension would allow for the generation of CMB maps directly from theoretical power spectra using QFT, thereby replacing the classical FFT in both forward and inverse transformations. Notably, data encoding may be less critical in this case, as the pipeline requires only \(C_\ell\) values, which scale as \textit{nside}, rather than the full set of \(a_{\ell m}\) coefficients, which scale as \(\textit{nside}^2\). This reduced data requirement could make quantum supremacy more feasible in this context. Furthermore, the generation of \(a_{\ell m}\) as random variables with variance \(C_\ell\), possibly following a Gaussian distribution, could further benefit from quantum computing, enabling exact and potentially faster quantum-based random number generation instead of classical pseudo-random methods.

Furthermore, a critical area for improvement lies in optimizing quantum data encoding. Future efforts will explore alternative encoding strategies or hybrid quantum-classical approaches to mitigate the significant overhead currently associated with amplitude encoding. Advancements in this direction could substantially enhance the efficiency of QFT-based spherical harmonic analysis, making it a viable alternative to FFT in CMB map-making and other astrophysical data processing applications.

\section*{Acknowledgment}
This research was supported by the ICSC Italian national research center on High Performance Computing, Big Data, and Quantum Computing. This work utilized IBM Quantum services and Qiskit for quantum simulations and execution. We acknowledge the contributions of the open-source quantum computing community. AI usage disclosure: ChatGPT-4 was employed for editing and grammar checks throughout the entire paper to enhance clarity and readability; as well as rephrasing sentences in Sections \ref{intro}, \ref{cmb}, \ref{map}, and \ref{qft}.


\bibliographystyle{unsrt}
\bibliography{ref}

\end{document}